\documentclass[12pt]{article}
\usepackage[english]{babel}
\usepackage[a4paper,top=2.54cm,bottom=2.54cm,left=3cm,right=3cm,marginparwidth=2.5cm]{geometry}
\usepackage{amsmath}
\usepackage{amssymb}
\usepackage{graphicx}
\usepackage{times}
\usepackage[colorlinks=true, allcolors=blue]{hyperref}
\usepackage[autostyle]{csquotes}
\usepackage[
    backend=biber,
    style=ieee,
    natbib=true,
    url=false, 
    doi=true,
    eprint=false,
    isbn=false,
    date=year,
    citestyle=numeric-comp
]{biblatex}
\usepackage{mmacells}
\mmaSet{leftmargin=0.0em}

\lstdefinelanguage{Ini}
{
    basicstyle=\ttfamily\small,
    columns=fullflexible,
    morecomment=[s][\bfseries]{[}{]},
    morecomment=[l]{\#},
    morecomment=[l]{;},
    commentstyle=\color{gray}\ttfamily,
    morekeywords={},
    otherkeywords={=,:},
    keywordstyle={\bfseries}
}

\newcommand{\rmi}{{\rm i}}
\newcommand{\rme}{{\rm e}}
\newcommand{\rmd}{{\rm d}}

\def\bra#1{\mbox{$\langle#1|$}}
\def\ket#1{\mbox{$|#1\rangle$}}

\title{Open quantum system simulation of time and frequency resolved spectroscopy}
\author{Tobias Kramer\footnote{tobias.kramer@jku.at}}
\date{\small Institute for theoretical physics\\
Department of quantum and classical dynamics\\
Johannes Kepler Universität Linz, Austria}
\begin{document}
\maketitle

\begin{abstract}
The dynamics of excitonic energy transfer in molecular complexes triggered by interaction with laser pulses offers a unique window into the underlying physical processes.
The absorbed energy moves through the network of interlinked pigments and in photosynthetic complexes reaches a reaction center.
The efficiency and time-scale depend not only on the excitonic couplings, but are also affected by the dissipation of energy to vibrational modes of the molecules.
An open quantum system description provides a suitable tool to describe the involved processes and connects the decoherence and relaxation dynamics to measurements of the time-dependent polarization.
\end{abstract}

\section{Introduction}

This lecture note reviews how to perform detailed calculations of the dynamics in open quantum systems with applications to energy transfer in light-harvesting complexes.
The approach relies on the Frenkel exciton description of excitonic energy transfer.
For an introduction to the Frenkel exciton picture from a molecular theory perspective, we refer to the monograph by May and Kühn \cite{May2004}, as well as other relevant materials from the Les Houches school on ``Quantum Dynamics and Spectroscopy of Functional Molecular Materials and Biological Photosystems'', in particular the contribution by Renger.

Sect.~2 reviews the open quantum system approach, and in Sect.~3 we compare different solution methods. Sect.~4 describes the computation of optical spectra, while Sect.~5 reviews the ensemble averages required to compare with experimental data.
As a well-studied light-harvesting complex with a known atomistic structure, we focus exemplary calculations in Sect.~6 on the Fenna-Matthews-Olson (FMO) complex. 
Its structure model, derived from experimental observations on crystallized complexes, is detailed in the seminal works by Olson et al. \cite{Olson1962} and Fenna and Matthews \cite{Fenna1975}.
Parameters for the corresponding Frenkel exciton model have been derived by Adolphs and Renger \cite{Adolphs2006}.

\section{Open quantum system dynamics}

To describe the dynamics of an electronic excitation in an LHC, the concept of an \textit{open quantum system} is employed.
A comprehensive introduction to open quantum system dynamics can be found in \cite{carmichael_statistical_1999}, and a concise overview of the notation applicable to LHC is provided in \cite{May2004}.
For the purpose of defining the fundamental quantities of interest, this section closely follows the approach outlined in \cite{Kramer2014,Kramer2018e}.
The photosynthetic complex, which interacts with light, is described using the Frenkel exciton model.
In this model, the system is characterized by the Hamiltonian given by Equation (\ref{eq:h-full}):
\begin{equation}
\label{eq:h-full}
H(t)=H_{\rm g}+H_{\rm ex}+H_{\rm bath}+H_{\rm ex-bath}+H_{\rm field}(t).
\end{equation}
The first term, denoted as $H_{\rm g}=\varepsilon_0\ket{0}\bra{0}$, represents the ground state Hamiltonian with a ground state energy of $\varepsilon_0$. 
The second term, $H_{\rm ex}$, incorporates the excitation energies of each pigment and the inter-pigment couplings.
Additionally, the vibrational modes of each pigment are introduced through the inclusion of $H_{\rm bath}$. In this model, the vibrations are linearly coupled to the excitons via $H_{\rm ex-bath}$.
When expressed in bra-ket notation, the excitonic Hamiltonian $H_{\rm ex}$ is formulated for a system comprising $N_{\rm sites}$ pigments (referred to as \textit{sites}) as follows in the site basis ($H_{\rm ex}^{\rm site}$):
\begin{equation}
\label{eq:h-ex} 	
H_{0}^{\rm site} =\sum_{m=1}^{N_{\rm sites}}\varepsilon_m^0\ket{m}\bra{m}+\sum_{n\neq m}J_{mn}\ket{m}\bra{n},\quad
H_{\rm ex}^{\rm site} =H_{0}^{\rm site}
+\sum_{m=1}^{N_{\rm sites}}\sum_{v=1}^{V_m}\lambda_{m,v}\ket{m}\bra{m}.
\end{equation}
The site energy $\varepsilon_{m}=\varepsilon_{m}^{0}+\sum_{v=1}^{V_m}\lambda_{m,v}$ consists of the zero phonon energy  $\varepsilon_{m}^{0}$ plus the reorganization energy $\sum_v\lambda_{m,v}$.
The inter-site couplings are denoted by $J_{mn}$.
The vibrational states of the pigments are described by $B=\sum_m V_m$  independent baths, where several baths can be assigned to the same pigment to either represent a more complex spectral density or to describe states representing two excitons at different sites.
Non-diagonal (site $m\ne n$) couplings between a state $\ket{m}\bra{n}$ and baths are not considered here.
The $V_m$ baths are represented by a  Hamiltonian  $H_{{\rm bath},m,v}=\sum\limits_{i}\hbar\omega_{m,v,i}(b_{m,v,i}^{\dagger}b_{m,v,i}+\frac{1}{2})$ which consist of harmonic oscillators with frequencies $\omega_{m,v,i}$.
The bosonic creation and annihilation operators $b_{m,v,i}$ are specified for each pigment $m$.
Finally, the linear coupling to the excitons is written in terms of the linear displacement of each bath mode $(b^{\dagger}_{m,v,i}+b_{m,v,i})$:
\begin{equation}
H_{\rm ex-bath}=\sum_m \ket{m}\bra{m} \otimes \sum_v\sum_i \hbar\omega_{m,v,i} d_{mvi}(b^{\dagger}_{m,v,i}+b_{m,v,i}),
\end{equation}
Here, $d_{mvi}$ denotes the coupling strength, which for a continuous spectrum of oscillator frequencies is expressed as spectral density of vibrational modes 
\begin{equation}
J_{m,v}(\omega)=\pi\sum_i \hbar^2\omega_{mvi}^2d_{mvi}^2\delta(\omega-\omega_{mvi}).
\end{equation}
The spectral density is also connected to the reorganization energy
\begin{equation}
\lambda_{m,v}=\int_0^\infty \frac{J_{m,v}(\omega)}{\pi\omega}\rmd\omega.
\end{equation}
The Liouville-von Neumann equation describes the dynamics of an open quantum system in terms of the full (system and bath) density matrix $\rho^{\rm total}(t)$:
\begin{equation}
\label{eq:LNE}
\frac{\partial}{\partial t}\rho^{\rm total}(t)=-\frac{\rmi}{\hbar}[H(t),{\rho}^{\rm total}(t)].
\end{equation}
For the description of the excitonic degrees-of-freedom and the optical response of the system to light pulses the vibrational degrees of freedom are traced out.
The remaining \textit{reduced density matrix} ${\rho}(t)$ becomes
\begin{equation}
\label{eq:reduced_density_matrix}
{\rho}(t)={\rm Tr}_{\rm bath}\left[{\rho}^{\rm total}(t)\right]. 
\end{equation}
The reduced density matrix is in general evolving in a non-unitary fashion, in contrast to the total density matrix.

\section{Exact vs.\ approximate solution}

The density matrix of an isolated system (i.e.\ without coupling to vibrations) undergoes a coherent dynamics.
Decoherence and relaxation is brought into the system dynamics by the specifics of the coupling to vibrational modes, which affects the system dynamics.
We use a parametrization of the vibrational modes introduced in \cite{Kreisbeck2012} and implemented in \cite{Kreisbeck2014} and \cite{Kramer2018e,Noack2018a}.
To describe a frequency-dependent vibrational bath we use a superposition of (shifted) Drude-Lorentz peaks assigned to each site:
\begin{equation}
\label{eq:spectral_density_DL}
J_{m}(\omega)=\sum_{v=1}^{V_m}
\left(
 \frac{\lambda_{m,v}\omega\nu_{m,v}}{{(\omega-\Omega_{m,v})}^2+\nu_{m,v}^2}
+\frac{\lambda_{m,v}\omega\nu_{m,v}}{{(\omega+\Omega_{m,v})}^2+\nu_{m,v}^2}\right).
\end{equation}
Here, $\nu_{m,v}^{-1}$ denotes the inverse bath correlation time and the parameter $\Omega_{m,v}$ shifts the peak position of the spectral density and allows one to vary the pure dephasing and relaxation processes, while maintaining the reorganization energy $\lambda_{m,v}$ \cite{Kreisbeck2012,Kramer2014}.

There are several methods available to evolve a reduced density matrix of a system linearly coupled to a bath as given by Eq.~(\ref{eq:h-ex}) to various degrees of approximation.
For arbitrary coupling $\lambda$, the solution requires numerical methods, such as the the Hierarchical Equation Of Motion (HEOM) \cite{Tanimura1989,Tanimura1994}.
The HEOM method serves as the standard reference for comparing with other methods, including the quasi-adiabatic path integral QUAPI \cite{makri_numerical_1995} and various stochastic methods.
A Mathematica implementation of HEOM is available online at \cite{kramer_mathematica_2023} as reference implementation for demonstrating the algorithm.
We discuss high-performance implementations of HEOM in Sec.~\ref{sec:heom}, which are suitable for computing quantum dynamics and optical response functions in larger systems (up to 100 sites).

\subsection{Weak coupling limit: Redfield equations}

To explore the dynamics of the coupled system, it is useful to consider first only a weak system-bath coupling (small $\lambda$ compared to the eigenenergy differences). 
In this limit, the Redfield equation provides a suitable approximation (a concise derivation can be found in \cite{Ishizaki2009,Ishizaki2009b}). 
When a system is coupled to a thermal environment, it eventually reaches an equilibrium state where both the system and the environment share a common temperature. 
The timescale for thermalization is inversely proportional to the relaxation rate. 
The \textit{Redfield approach} reveals that, in the case of weak coupling, the relaxation rate depends on the spectral density value at the energy difference between two energy eigenstates.
Under weak coupling conditions the system and environment can be considered as forming a product state, with entanglement considered as negligible perturbation (Born approximation).

\begin{figure}
   \begin{center}
      \includegraphics[width=0.75\textwidth]{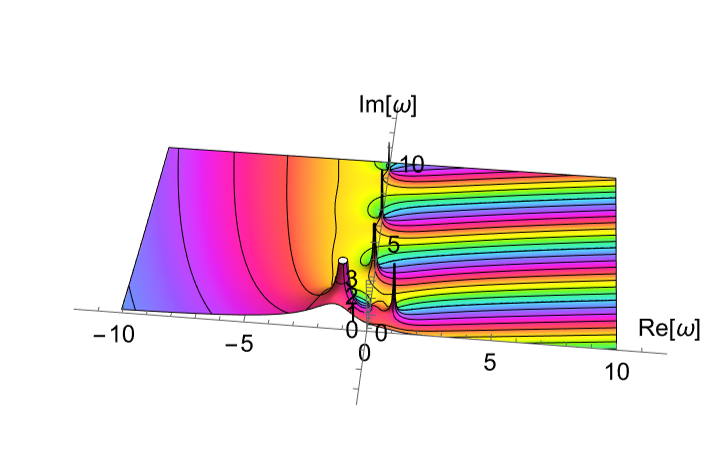}
   \end{center}
   \caption{\label{fig:complexplane}
   Visualization of the integrand $c(\omega)=J(\omega)n(\omega,\beta)\rme^{\rmi\omega t}$ of $C(t)=\int_{-\infty}^\infty c(\omega) \rmd\omega $ in the complex $\omega$ plane in the domain $-10<\Re\omega<10$ and $0<\Im\omega<10$.
   To obtain $C(t)$, the integration path along the real $\omega$-axis is deformed and closed by a semi-circular contour in the positive imaginary $\omega$-plane.
   This new path encloses an infinite number of poles, which yield the result of the integration by a sum of the residues at the poles, multiplied by $2\pi\rmi$.} 
\end{figure}

The Redfield tensor $R$ is commonly expressed in the energy representation, which is connected to the site representation through the diagonalizing matrix $A$:
\begin{equation}
H^{\rm exc}={A} H^{\rm site} {A}^T.
\end{equation}
In the energy basis, the matrix $H^{\rm exc}$ only has diagonal entries, with $i=1,\ldots,N_{\rm sites}$ representing the eigenenergies $E_i=\hbar\omega_i$.
The Redfield tensor is then entirely determined by the (in general site dependent) bath correlation function $C_m(\omega)$, which in turn depends on the spectral density (as described below):
\begin{eqnarray}
R_{\mu\nu\mu'\nu'}&=&\Gamma_{\mu\nu\mu'\nu'}+
{(\Gamma_{\mu\nu\mu'\nu'})}^* -\delta_{\nu\nu'} \sum_{\kappa=1}^{N_{\rm sites}}\Gamma_{\mu\kappa\kappa\mu'}
-\delta_{\mu\mu'} \sum_{\kappa=1}^{N_{\rm sites}}\Gamma_{\nu\kappa\kappa\nu'},\\
\Gamma_{\mu\nu\mu'\nu'}&=&\frac{1}{\hbar^2}
\sum_{m=1}^{N_{\rm sites}}A_{\mu m}A_{\nu m}A_{\mu' m}A_{\nu' m}
C_m(\omega_{\nu'}-\omega_{\mu'}), 
\end{eqnarray}
The Fourier transform of the bath correlation function at temperature $T$ ($\beta=1/(k_B T)$) is given by
\begin{eqnarray}
C_m(t)&=&\frac{1}{\pi} \int_0^\infty\rmd\omega\;J_m(\omega) \left[n(\omega,\beta)\rme^{\rmi\omega t}+(n(\omega,\beta)+1)\rme^{-\rmi\omega t}\right]\\
&=&\frac{1}{\pi} \int_{-\infty}^\infty\rmd\omega\;J_m(\omega) n(\omega,\beta)\rme^{\rmi\omega t}, \quad n(\omega,\beta)=1/(\rme^{\beta\hbar\omega}-1).\label{eq:cmt}
\end{eqnarray}
The integrand of Eq.~(\ref{eq:cmt}) is depicted in 
Fig.~\ref{fig:complexplane}.
For the spectral density in Eq.~(\ref{eq:spectral_density_DL}) it is expressed in terms of the Digamma function $\digamma$ (we surppress the index $m$ for compactness):
{\small 
\begin{eqnarray}
C(\omega)&=&-\frac{\rmi \lambda  \hbar}{2} \bigg[\frac{\nu _+ \left(-\omega +\rmi \nu _-\right) \cot \left(\frac{1}{2} \beta  \nu _+ \hbar \right)+\rmi \nu _- \left(\nu +\rmi
   \omega _+\right) \cot \left(\frac{1}{2} \beta  \nu _- \hbar \right)-2 \left(\nu ^2+\rmi \nu  \omega +\Omega ^2\right)}{\Omega ^2+(\nu +\rmi \omega )^2}\nonumber \\
&& +\frac{\rmi
   \nu _+ \left(\nu ^2+\omega _+^2\right) \digamma \left(\frac{\beta  \nu _+ \hbar }{2 \pi }+1\right)}{\pi  \left(\nu -\rmi \omega _-\right) \left(\nu ^2+\omega _+^2\right)}\\\nonumber
&& +\frac{\rmi \nu _- \left(\nu _+^2+\omega ^2\right) \digamma
   \left(\frac{\beta  \nu _- \hbar }{2 \pi }+1\right)-\rmi \nu _+ \left(\Omega ^2+(\nu -\rmi \omega )^2\right) \digamma \left(1-\frac{\beta  \nu _+ \hbar }{2 \pi
   }\right)}{\pi  \left(\nu -\rmi \omega _-\right) \left(\nu ^2+\omega _+^2\right)}\\\nonumber
&&+\frac{2 \nu  \omega  \left(\frac{1}{\nu ^2+\omega _+^2}+\frac{1}{\nu ^2+\omega
   _-^2}\right) \digamma \left(1+\frac{\rmi \beta  \omega  \hbar }{2 \pi }\right)}{\pi }+\frac{\nu _- \digamma \left(1-\frac{\beta  \nu _- \hbar }{2 \pi
   }\right)}{\pi  \left(-\omega +\rmi \nu _-\right)}\bigg],\\
\nu_\pm&=&\nu\pm\rmi \Omega\\
\omega_\pm&=&\omega\pm\rmi \Omega
\end{eqnarray}}
~\\[-1ex]
The last relation can be computed by using the residues theorem for $C(t)$ and by performing the Fourier transformation $(t \rightarrow \omega)$ term by term.
The Redfield tensor is then evaluated in terms of $C(\omega)$ values and the time evolution of the density matrix elements $\rho_{\mu\nu}$ in the energy representation of the exciton Hamiltonian (\ref{eq:h-ex}) is a solution to the following differential \textit{full Redfield equation}:
\begin{equation}
\label{eq:Redfield}
\frac{\partial \rho^{\rm exc}_{\mu\nu}(t)}{\partial t} = -\rmi (\omega_{\mu}-\omega_{\nu})\rho^{\rm exc}_{\mu\nu}(t) + \sum_{\mu'=1}^{N_{\rm states}}\sum_{\nu'=1}^{N_{\rm states}}R_{\mu\nu,\mu'\nu'}\rho^{\rm exc}_{\mu'\nu'}(t).
\end{equation}
The first term in eq.~(\ref{eq:Redfield}) represents the coherent evolution governed by the diagonalized Hamiltonian, while the second term accounts for decoherence and relaxation resulting from the coupling to the baths.
To simplify the equations further, we can make use of the \textit{secular Redfield} approximation. In this approximation, all entries that do not satisfy the condition $(\omega_\mu-\omega_\nu)=(\omega_{\mu'}-\omega_{\nu'})$ are set to zero. This approximation is employed to address the violation of positive definiteness in the density matrix that can occur at low temperatures when using the full Redfield equations. 
A detailed comparison of the dynamics obtained using the full and secular Redfield equations is presented in \cite{Kramer2018e}.

To compare the results with the reduced density matrix obtained from HEOM in the site basis, we need to transform the Redfield density-matrix from the energy basis back to the site representation:
\begin{equation}
\rho_{\rm Redfield}^{\text{site}}(t) = {A}^T \rho_{\rm Redfield}^{\rm exc}(t) {A}.
\end{equation}

\subsection{Strong coupling limit: Förster energy transfer and rate equations}

The Redfield description relies on a weak coupling between system and bath ($\lambda$ small compared to eigenenergy differences).
In the opposite case, for a very strong coupling, Förster theory provides an alternative approach to compute the quantum dynamics.
The F\"orster expression for the rate $\mathbf{R}$ in site basis reads \cite{Yang2002}
\begin{equation}
R_{m,n}^\text{Förster}=2|J_{mn}| \;\Re\! \left[\int_0^\infty \rmd t F_m^*(t) A_n(t)\right],
\end{equation}
with 
\begin{eqnarray}
A_n  (t)&=&\exp[-\rmi (\epsilon_n^0+\lambda_n) t-g_n(t)],\\
F_m^*(t)&=&\exp[+\rmi (\epsilon_m^0-\lambda_m) t-g_m(t)],\\
g_m(t)&=&-\frac{1}{2\pi}\int_{-\infty}^\infty\rmd\omega \; \frac{J_m(\omega)}{\omega^2}
\left(1+\coth(\beta\hbar\omega/2)\right)\left(\rme^{-\rmi\omega t}+\rmi\omega t-1\right).
\end{eqnarray}
The Förster rate is therefore determined by the overlap of absorption ($A$) and emission ($F$) spectra of the monomers, computed using the lineshape function $g(t)$ \cite{kubo_application_1955}.
The lineshape function is in turn given as double integral of the bath correlation-function $C(t)$ \cite{knox_low-temperature_2002}
\begin{equation}
    g(t)=\int_0^t\rmd\tau_1\int_0^{\tau_1}  \rmd\tau_2 \, C(\tau_2).
\end{equation}
The last relation connects the spectral density of a monomeric unit to the absorption spectrum at very low temperatures via the lineshape function.
An experimental determination of the spectral density of the FMO complex from fluorescence line-narrowing measurements is performed in \cite{wendling_electron-vibrational_2000}.
The population dynamics in F\"orster theory is then given by
\begin{eqnarray}
\rho_{mm}(t)&=&\rho_{mm}(0) \rme^{\mathbf{K} t},\\
K_{\alpha\alpha}&=&-\sum_{\gamma=1,\gamma\ne\alpha}^{N} R_{\gamma\alpha}^\text{Förster},\\
K_{\alpha\beta}&=&R_{\alpha\beta}^\text{Förster}, \quad (\alpha\ne\beta)
\end{eqnarray}

For the shifted Drude Lorentz spectral density (\ref{eq:spectral_density_DL}), an analytic expression for the bath-correlation and lineshape function can be computed using \textit{Mathematica}:
{\footnotesize
\begin{mmaCell}[morelst={breaklines=true},moredefined={g},morepattern={t_}]{Input}
      g[t_]=\mmaFrac{\mmaSup{\mmaDef{e}}{-t \mmaUnd{\(\pmb{\nu}\)}} \mmaUnd{\(\pmb{\lambda}\)} (\mmaSup{\mmaDef{e}}{\mmaDef{i} t \mmaUnd{\(\pmb{\Omega}\)}}+\mmaSup{\mmaDef{e}}{t
    \mmaUnd{\(\pmb{\nu}\)}} (-1+t (\mmaUnd{\(\pmb{\nu}\)}-\mmaDef{i} \mmaUnd{\(\pmb{\Omega}\)}))) (-\mmaDef{i}+Cot[\mmaFrac{(\mmaUnd{\(\pmb{\nu}\)}-\mmaDef{i}
    \mmaUnd{\(\pmb{\Omega}\)}) \mmaUnd{\(\pmb{\hbar}\)}}{2 kB T}])}{2 (\mmaUnd{\(\pmb{\nu}\)}-\mmaDef{i} \mmaUnd{\(\pmb{\Omega}\)})}+\mmaFrac{\mmaSup{\mmaDef{e}}{-t
    (\mmaUnd{\(\pmb{\nu}\)}+\mmaDef{i} \mmaUnd{\(\pmb{\Omega}\)})} \mmaUnd{\(\pmb{\lambda}\)} (-\mmaDef{i}+\mmaSup{\mmaDef{e}}{t (\mmaUnd{\(\pmb{\nu}\)}+\mmaDef{i}
    \mmaUnd{\(\pmb{\Omega}\)})} (\mmaDef{i}+t (-\mmaDef{i} \mmaUnd{\(\pmb{\nu}\)}+\mmaUnd{\(\pmb{\Omega}\)}))) (1+Coth[\mmaFrac{(-\mmaDef{i} \mmaUnd{\(\pmb{\nu}\)}+\mmaUnd{\(\pmb{\Omega}\)})
    \mmaUnd{\(\pmb{\hbar}\)}}{2 kB T}])}{2 (\mmaUnd{\(\pmb{\nu}\)}+\mmaDef{i} \mmaUnd{\(\pmb{\Omega}\)})}+\mmaFrac{1}{2 \mmaDef{\(\pmb{\pi}\)}}\mmaUnd{\(\pmb{\lambda}\)}
    (\mmaFrac{(1-t \mmaUnd{\(\pmb{\nu}\)}+\mmaDef{i} t \mmaUnd{\(\pmb{\Omega}\)}) HarmonicNumber[-\mmaFrac{(\mmaUnd{\(\pmb{\nu}\)}-\mmaDef{i} \mmaUnd{\(\pmb{\Omega}\)})
    \mmaUnd{\(\pmb{\hbar}\)}}{2 kB \mmaDef{\(\pmb{\pi}\)} T}]}{\mmaUnd{\(\pmb{\nu}\)}-\mmaDef{i} \mmaUnd{\(\pmb{\Omega}\)}}+(t+\mmaFrac{1}{\mmaUnd{\(\pmb{\nu}\)}-\mmaDef{i}
    \mmaUnd{\(\pmb{\Omega}\)}}) HarmonicNumber[\mmaFrac{(\mmaUnd{\(\pmb{\nu}\)}-\mmaDef{i} \mmaUnd{\(\pmb{\Omega}\)}) \mmaUnd{\(\pmb{\hbar}\)}}{2 kB
    \mmaDef{\(\pmb{\pi}\)} T}]+\mmaFrac{(1-t (\mmaUnd{\(\pmb{\nu}\)}+\mmaDef{i} \mmaUnd{\(\pmb{\Omega}\)})) HarmonicNumber[-\mmaFrac{(\mmaUnd{\(\pmb{\nu}\)}+\mmaDef{i}
    \mmaUnd{\(\pmb{\Omega}\)}) \mmaUnd{\(\pmb{\hbar}\)}}{2 kB \mmaDef{\(\pmb{\pi}\)} T}]}{\mmaUnd{\(\pmb{\nu}\)}+\mmaDef{i} \mmaUnd{\(\pmb{\Omega}\)}}+\mmaFrac{(1+t
    \mmaUnd{\(\pmb{\nu}\)}+\mmaDef{i} t \mmaUnd{\(\pmb{\Omega}\)}) HarmonicNumber[\mmaFrac{(\mmaUnd{\(\pmb{\nu}\)}+\mmaDef{i} \mmaUnd{\(\pmb{\Omega}\)})
    \mmaUnd{\(\pmb{\hbar}\)}}{2 kB \mmaDef{\(\pmb{\pi}\)} T}]}{\mmaUnd{\(\pmb{\nu}\)}+\mmaDef{i} \mmaUnd{\(\pmb{\Omega}\)}}+\mmaFrac{1}{\mmaSup{\mmaUnd{\(\pmb{\nu}\)}}{2}+\mmaSup{\mmaUnd{\(\pmb{\Omega}\)}}{2}}\mmaSup{\mmaDef{e}}{-\mmaFrac{2
    kB \mmaDef{\(\pmb{\pi}\)} t T}{\mmaUnd{\(\pmb{\hbar}\)}}} ((\mmaUnd{\(\pmb{\nu}\)}+\mmaDef{i} \mmaUnd{\(\pmb{\Omega}\)}) (HurwitzLerchPhi[\mmaSup{\mmaDef{e}}{-\mmaFrac{2
    kB \mmaDef{\(\pmb{\pi}\)} t T}{\mmaUnd{\(\pmb{\hbar}\)}}},1,1-\mmaFrac{(\mmaUnd{\(\pmb{\nu}\)}-\mmaDef{i} \mmaUnd{\(\pmb{\Omega}\)}) \mmaUnd{\(\pmb{\hbar}\)}}{2
    kB \mmaDef{\(\pmb{\pi}\)} T}]+HurwitzLerchPhi[\mmaSup{\mmaDef{e}}{-\mmaFrac{2 kB \mmaDef{\(\pmb{\pi}\)} t T}{\mmaUnd{\(\pmb{\hbar}\)}}},1,1+\mmaFrac{(\mmaUnd{\(\pmb{\nu}\)}-\mmaDef{i}
    \mmaUnd{\(\pmb{\Omega}\)}) \mmaUnd{\(\pmb{\hbar}\)}}{2 kB \mmaDef{\(\pmb{\pi}\)} T}])+(\mmaUnd{\(\pmb{\nu}\)}-\mmaDef{i} \mmaUnd{\(\pmb{\Omega}\)})
    (HurwitzLerchPhi[\mmaSup{\mmaDef{e}}{-\mmaFrac{2 kB \mmaDef{\(\pmb{\pi}\)} t T}{\mmaUnd{\(\pmb{\hbar}\)}}},1,1-\mmaFrac{(\mmaUnd{\(\pmb{\nu}\)}+\mmaDef{i}
    \mmaUnd{\(\pmb{\Omega}\)}) \mmaUnd{\(\pmb{\hbar}\)}}{2 kB \mmaDef{\(\pmb{\pi}\)} T}]+HurwitzLerchPhi[\mmaSup{\mmaDef{e}}{-\mmaFrac{2 kB \mmaDef{\(\pmb{\pi}\)}
    t T}{\mmaUnd{\(\pmb{\hbar}\)}}},1,1+\mmaFrac{(\mmaUnd{\(\pmb{\nu}\)}+\mmaDef{i} \mmaUnd{\(\pmb{\Omega}\)}) \mmaUnd{\(\pmb{\hbar}\)}}{2 kB \mmaDef{\(\pmb{\pi}\)}
    T}])+4 \mmaSup{\mmaDef{e}}{\mmaFrac{2 kB \mmaDef{\(\pmb{\pi}\)} t T}{\mmaUnd{\(\pmb{\hbar}\)}}} \mmaUnd{\(\pmb{\nu}\)} Log[1-\mmaSup{\mmaDef{e}}{-\mmaFrac{2
    kB \mmaDef{\(\pmb{\pi}\)} t T}{\mmaUnd{\(\pmb{\hbar}\)}}}]))
\end{mmaCell}
}

While Redfield and Förster theories correspond respectively to the limiting cases of weak or strong coupling of the system to the thermal environment, the HEOM method developed by Kubo and Tanimura \cite{Tanimura1989} provides the  connection between both regimes and in addition covers both limits \cite{Ishizaki2009b}.
Within HEOM, the time evolution of the reduced density matrix is described by a system of interlinked differential equations comprising  $N_{\rm matrices}$ auxiliary density matrices $\sigma_u$ of dimensions $N_\text{states}\times N_\text{states}$.
The auxiliary density matrices (also referred to \textit{auxiliary density operators (ADO)}) are put in separate layers with a specific depth index $D$.
The matrices in different layers are connected by vertices with $+$-upward and $-$-downward connecting links. 
\begin{equation}
\frac{\rm d \sigma_u}{\rmd t}
=-\frac{\rmi}{\hbar}\left[H, \sigma_u\right]
+\sum_{\rm baths} A \sigma_{u}
+\sum_{\rm baths} B \sigma_{u_{+}}
+\sum_{\rm baths} C \sigma_{u_{-}}.
\end{equation}
Explicit expressions for the operators $A,B,C$ are given in 
\cite{Tanimura2006} and \cite{Kramer2018e}, Eqs.~(12-36).
The number of matrices in each layer increases with the layer depth $D$, whereas the top-layer contains as unique member the reduced density matrix $\sigma_0$.
The layer links are the result of a series expansion of the exponentially decaying bath correlation function and thus contains with increasing depth increasingly higher derivatives.
The HEOM depth has to be carefully chosen to guarantee a converged result.
For the parameters encountered in the FMO complex this implies that only the first few layers are required ($D=2$-$3$) at $T=300$~K \cite{Kramer2018e}.
For lower temperatures or a stronger system-bath coupling, the total number of matrixes increases and is given by
\begin{equation}
   \label{eq:Nmatrices}
   N_{\rm matrices}=
   \left(\begin{array}{c}
   M B+D\\
   M B
   \end{array}\right),
\end{equation}
where $B$ denotes the number of vibrational baths $B$, and $M$ the number of Matsubara modes $M$ \cite{Kramer2018e}.
In practise this large number of matrices limits the HEOM method for computing exciton dynamics to systems with less than 100~pigments.
A detailed comparison of Förster theory with HEOM in the Photosystem~I complex \cite{Kramer2018d} shows that the aggregated transfer times from the A-B branch within the Förster theory (9~ps) differ from the exact HEOM results in the presence of a mixture of small and large intersite couplings (17~ps).

\begin{figure}
\centering
\includegraphics[width=0.5\textwidth]{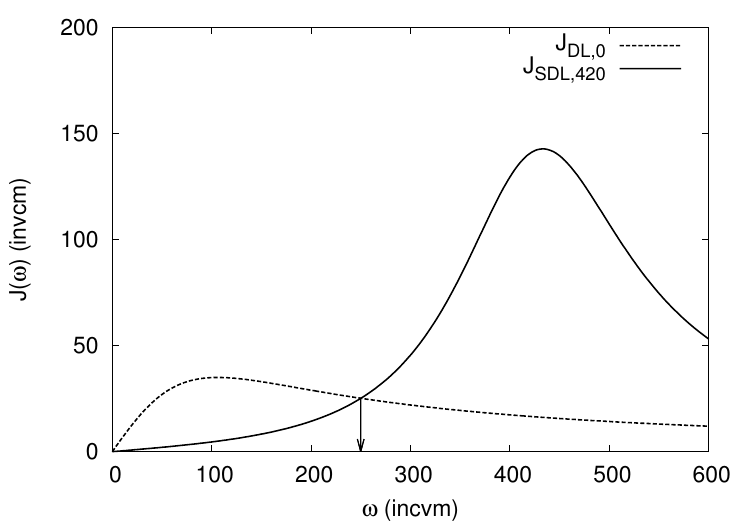}
\includegraphics[width=0.4\textwidth]{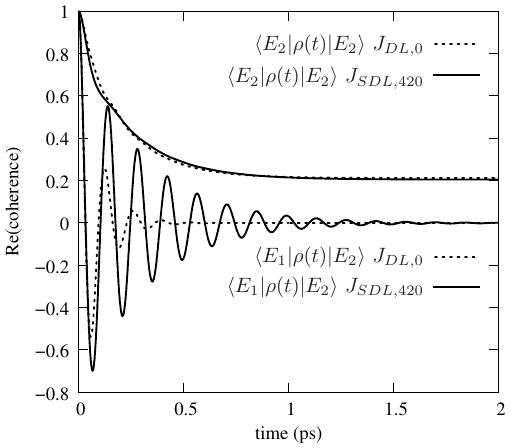}
\caption{\label{fig:coherence}
Left panel:
Spectral density $J_{\rm DL,0}$ (unshifted Drude-Lorentz form, $\lambda=35$~cm$^{-1}$ and $\nu^{-1}=50$~fs) and $J_{\rm SDL,420}$ 
(shifted Drude-Lorentz peak, $\Omega=420$~cm$^{-1}$, $\lambda=35$~cm$^{-1}$ and $\nu^{-1}=50$~fs).
The arrow indicates the difference of eigenenergies of a two-site system 
$
H_{\rm ex}=\left(
\begin{array}{cc}
-75 & 100 \\
100 & 75 
\end{array}
\right)\;{\rm cm}^{-1}$, where by construction both spectral densities have the same value.
Right panel:
Relaxation of the diagonal element $\langle E_2| \rho(t) |E_2 \rangle$ to the thermal state (upper non-oscillatory graphs) and damped oscillations of the off-diagonal coherence ${\rm Re} (\langle E_1| \rho(t) |E_2 \rangle)$ at $T=277$~K.
While both spectral densities give very similar relaxation rates, the off-diagonal coherence is much prolonged for $J_{\rm SDL,420}$ due to its small slope toward $\omega\rightarrow 0$.
Reprinted from \cite{Kramer2014}, Fig.~8, with the permission of AIP Publishing.}
\end{figure}

\begin{figure}[t]
\centering
\includegraphics[width=0.7\textwidth]{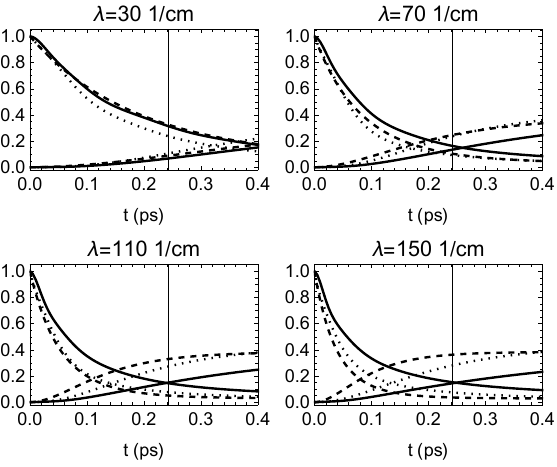}
\caption{\label{fig:themalization} Quantum dynamics starting from the highest eigenstate of the FMO Hamiltonian \cite{Adolphs2006} to the thermal state for various reorganization energies $\lambda$ at $T=277$~K.
Solid line: HEOM method (exact) population in exciton basis of the highest and lowest eigenstate populations, dashed line: secular Redfield theory, dotted line: full Redfield.
HEOM shows there exists an optimal value of $\lambda\approx 110$~cm${}^{-1}$ for fastest thermalization (as seen by the crossing of the populations of both states, vertical line), while in the secular Redfield approximation (not applicable at strong couplings) a higher coupling always increases the thermalization rate.
}
\end{figure}

\subsection{Decoherence, dephasing, and relaxation dynamics.}\label{sec:dec}

Any initial coherence, which is expressed by an off-diagonal element of the density matrix in the energy representation, decays over a timescale determined by the \textit{decoherence rate} $\gamma_\text{decoh}$.
The decoherence rate is set by two contributions $\gamma_\text{decoh}=\gamma_\text{pd}+\gamma_\text{r}/2$, the \textit{pure dephasing rate} $\gamma_\text{pd}$ determined by the slope of the spectral density as it approaches $\omega\rightarrow 0$ and the \textit{relaxation rate} $\gamma_\text{r}$, given by the value of the spectral density at the difference of eigenenergies.
For a two-site system, where each site is coupled to an independent bath at temperature $T$,
\begin{equation}
   H_{\rm exciton}=
   \begin{pmatrix}
   -\epsilon/2 & d/2 \\
   d/2 & \epsilon/2
   \end{pmatrix},
\end{equation}
the respective rates are (\cite{weiss_quantum_2012}, Sect.~21.4.2):
\begin{eqnarray*}
   \gamma_\text{r}&\approx& \frac{d^2        J(\omega)\coth(\hbar\omega/(2k_B T))}{2 (\epsilon^2+d^2)}\bigg|_{\omega=\sqrt{\epsilon^2+d^2}},\\
   \gamma_\text{pd}&=     & \frac{\epsilon^2 J(\omega)\coth(\hbar\omega/(2k_B T))}{2 (\epsilon^2+d^2)}\bigg|_{\omega\rightarrow 0}.
\end{eqnarray*}
An illustrative example is provided by a dimer which is coupled to an environment.
By choosing the form of the spectral density one can fix the thermalization rate, while at the same time the dephasing rate can be vastly different.
This is demonstrated in Fig.~\ref{fig:coherence}, reprinted from \cite{Kramer2014}.
For computing transport, the thermalization rate sets the time-scale of how fast energy is transferred towards a thermal equilibrium state. 
Secular Redfield theory would predict a faster equilibration for a stronger coupling to the vibrational states (i.e.\ for increasing $\lambda$), while the non-perturbative methods (i.e.\ HEOM) show that there exists an optimal value for $\lambda$ which supports the fastest thermalization (see \cite{Plenio2008,Rebentrost2009,Novoderezhkin2011a,Kreisbeck2011}).
Increasing $\lambda$ prolongs the thermalization process again, as illustrated in Fig.~\ref{fig:themalization}.
For comparison, also the full Redfield result is shown.
The good agreement between \textit{secular Redfield} and HEOM obtained for small reorganization energies relies on error compensation effects.
Modified Redfield theory, by the inclusion of multi-phonon relaxation processes, is also able to describe this effect via rate equations,  see, e.g.\ Fig.~2 in \cite{Yang2002}, but does not capture the dynamics of the coherences.

\section{Computing optical spectra}

The spectral density of the vibrational modes $J(\omega)$ directly influences the line widths computed for optical spectra.
To model the creation of an exciton through optical excitation of the molecular complex, we introduce the interaction between an external electric field and the dipole moments of the molecules. This interaction is represented by the dipole operator \cite{Chen2015}:
\begin{equation}
H_{\rm field}(t) = -\sum_p {\mathbf e_p} \cdot \hat{\mathbf \mu} E_p(\mathbf{r},t).
\end{equation}
Here, ${\mathbf e_p}$ is the unit vector in the Cartesian electric field component $E_p(\mathbf{r},t)$, and the dipole matrix operator is given by $\hat{\mathbf \mu}= \hat{\mathbf \mu}^++\hat{\mathbf \mu}^-$, where
\begin{equation}
\hat\mu^+=\sum_{a=1}^{N_{\rm sites}} \mathbf{d}_a |a\rangle\langle 0|,
\label{eq:mu_plus}
\end{equation}
\begin{equation}
\hat\mu^-=\sum_{a=1}^{N_{\rm sites}} \mathbf{d}_a|0\rangle\langle a|,=(\hat\mu^+)^\dagger.
\label{eq:mu_minus}
\end{equation}
To facilitate calculations, we decompose the real part of the time-varying electric field amplitude $E(\mathbf{r},t)=E^+(\mathbf{r},t)+ E^-(\mathbf{r},t)$ into two complex amplitudes, where $E_p^-(\mathbf{r},t)=(E_p^+(\mathbf{r},t))^*$ and
\begin{equation}
\label{eq:probe_pulse}
E^+(\mathbf{r},t)=\tilde{E}(t-t_{c})\rme^{\rmi(\omega_c t-\mathbf{k}\cdot\mathbf{r})}.
\end{equation}
In this expression, $\tilde{E}(t)$ represents the pulse envelope centered around time $t_{\rm c}$, $\omega_c$ is the carrier frequency, and $\varphi=\mathbf{k}\cdot\mathbf{r}$ denotes the phase of the laser pulse.
Using the rotating-wave approximation (RWA), the complex-valued electric field is combined with the respective excitation and de-excitation parts of the dipole operator \cite{Gelin2013,Chen2015},
reflecting energy conservation, that is, the excitation/de-excitation of the system is related to an annihilation/creation of a photon:
\begin{equation}
\label{eq:h-field-rwa}
H_{\rm field}(t) = -\sum_p {\mathbf e_p}\cdot [\hat{\mathbf \mu}^+ E_p^-(\mathbf{r},t) +\hat{\mathbf \mu}^- E_p^+(\mathbf{r},t)].
\end{equation}
To obtain the optical spectra, we examine the time-dependent optical response of the molecular complex, specifically the non-linear polarization $P(t)$ induced by a single or a combination of weak probe laser pulses. The time-dependent polarization is given by:
\begin{eqnarray}\label{eq:polarization}
P(t) 
&=& {\rm Tr}_\text{system}({\rm Tr}_\text{bath}({\rho}_\text{total}(t)) \hat{\mu}^+), \quad \rho_\text{total}(t=0)=|0\rangle\langle 0| \otimes \rho_\text{bath} \\
&=& {\rm Tr}_\text{system}(\rho(t) \hat{\mu}^+), \,\;\quad\qquad\qquad \rho(t)={\rm Tr}_\text{bath}({\rho}_\text{total}(t)),
\end{eqnarray}
where $\rho(t)$ represents the time-evolved reduced density matrix following the time-dependent Hamiltonian (\ref{eq:h-full}). 
For weak laser pulses, the polarization function can be expanded in powers of the electric field \cite{Hamm2011} and written as a convolution of the electric field with the response function $S^{(n)}(t_n,..,t_1)$ or calculated using a non-perturbative approach.

A spectrometer records for a linear absorption spectrum the sum of the incoming electric field of the laser $E_0$ and the polarization induced field $E_\text{signal}$ via its absolute value after performing a Fourier transform:
\begin{equation}
    I(\omega)={\left|\int_0^\infty\rmd t \rme^{\rmi \omega t} (E_0(t)+E_\text{signal}(t))\right|}^2\propto I_0(\omega)+2\Re \left[E_0(\omega) E_\text{signal}(\omega)\right].
\end{equation}
This expression neglects the weaker quadratic term due to the signal alone.
To compute the spectra, the dipole operator, which accounts for the charge redistribution in the presence of an external electric field in each molecule of the complex, must be specified. 
For short pulses, it is a $(N_{\rm sites}+1)\times (N_{\rm sites}+1)$ dimensional matrix, as shown in Eq.~(\ref{eq:mu_plus}).
For a specific cartesian component $p$, it reads:
\begin{equation}
\hat\mu^+_p=\sum_{m=1}^{N_{\rm sites}} \mathbf{e}_p \cdot \mathbf{d}_m |m\rangle\langle 0|.
\label{eq:mu_plusp}
\end{equation}
For long enough pulses or multiple short pulses, it is possible to excite an additional exciton, necessitating the extension of the dipole representation to include two-exciton states.
This enlarges the Hamiltonian and dipole matrix to $N_{\rm states}$ entries \cite{Cho2005,Hein2012}:
\begin{equation}
N_{\rm states}=1+ N_{\rm sites} + \left[N_{\rm sites} (N_{\rm sites} - 1)\right]/2.
\end{equation}

\subsection{Linear absorption spectra}
\label{subsec:absorption}

To compute the optical spectra, it is useful to employ the Fourier transform of the time evolution of the dipole correlation function, as described by \cite{Gordon1968} and earlier references therein. The Fourier transform allows us to obtain frequency-dependent spectra from the time-dependent trace.

First, we consider the linear absorption spectra, which result from a short initial excitation and can be obtained by Fourier transforming the dipole-dipole correlation function. 
To account for rotational averaging, we include the sum over polarization directions ${\mathbf e_p}$ (as discussed in Sect.~\ref{sec:rotav}). 
The linear absorption spectra $\langle{\rm LA}(\omega)\rangle_{\rm rot}$ is then expressed as:
\begin{equation}
\langle{\rm LA}(\omega)\rangle_{\rm rot} = {\rm Re}\sum_p\int^{\infty}_{0}\rmd t \exp(\rmi \omega t){\rm Tr}[\hat\mu_p(t)\hat\mu_p(0){\rho}(0)].
\label{eq:lin_abs_mukamel}
\end{equation}
Here, the dipole operators $\hat\mu_p(t)$ are calculated in the interaction picture \cite{Hamm2011}, and the trace is taken over the system part only, as the trace over the environment is already considered in the reduced density matrix.
At finite temperature, the presence of decoherence and relaxation towards the thermal state results in a decay of the time-dependent polarization, which, in the frequency domain, corresponds to a finite line-width. 
After the polarization vanishes, one can increase the range of time propagation by padding the time-series of the polarization with zeroes to longer times. 
This increases the sampling of the frequency-resolved spectra.

\subsection{Two-dimensional electronic spectroscopy (2DES)}\label{subsec:2d}

While the pure dephasing rate $\gamma_{\rm pd}$ determines the width of a single spectral line in the absorption spectra, the thermalization rate  $\gamma_{\rm r}$ determines the time-scale towards equilibrium of an intially excited state.
Two-dimensional electronic spectroscopy (2DES) allows one to track the excitonic energy transfer and to investigate the energetic arrangement of site (pigment) energies.

This is achieved by a pump-probe setup involving in total four pulses (including the signal pulse) and by scanning over a central interval (the time delay $T_2$) to observe the energy transfer.
Formally, 2DES reveals the third order non-linear response function, involving three commutators of the dipole operator acting at specific times $t_0=0, t_1, t_2$ and a fourth dipole operator acting at $t_3$ representing the signal.
In between the dipole operations, the reduced density matrix is propagated from $t_0=0$ to $t_3$ in $N_\text{steps}$.
The time traces of the response function in the first $T_1$ and last $T_3$ interval around the central interval are transformed to the frequency domain with two Fourier transforms $T_1 \rightarrow \omega_1$,
$T_3 \rightarrow \omega_3$ \cite{Mukamel1995,Hamm2011}.
To obtain the 2DES in the $\omega_1$-$\omega_3$ plane requires to obtain first a time-dependent data set in the $t_1, t_2$-plane for all times $t_i=i \Delta t$, with $i=0,\ldots,N_\text{steps}$ and time-step $\Delta t$.
The computation is performed for equidistantly spaced times $T_1=0,\Delta t,\ldots,t_1$ and $T_3=0,\Delta t,\ldots,(t_3-t_2)$. 
This requirement increases the computational overhead by a factor $N_\text{steps}$ compared to the calculation of absorption spectra.
In the impulsive limit, the 2D spectra can be separated by an expansion in terms products of the electric field of the different pulses with varying $\mathbf{k}$ vectors.
A separation into rephasing ($-\mathbf{k}_1+\mathbf{k}_2+\mathbf{k}_3$) and non-rephasing directions ($+\mathbf{k}_1-\mathbf{k}_2+\mathbf{k}_3$) results in three rephasing pathways and three non-rephasing pathways 
\cite{Mukamel1995,Hamm2011} representing ground state bleaching (GB), stimulated emission (SE), and excited state absorption (ESA).
In terms of the dipole operators at distinct times, the rephasing pathways are given by
\begin{eqnarray}
S_\text{GB,RP} (T_3,T_2,T_1|p_0,p_1,p_2,p_3)&=&+\rmi\,{\rm Tr}\big[{\hat\mu}_{p_3}^-(t_3){\hat\mu}_{p_2}^+(t_2)\rho_0{\hat\mu}_{p_0}^-(0){\hat\mu}_{p_1}^+(t_1)\big]\label{eq:2DESGBRP}\\
S_\text{SE,RP} (T_3,T_2,T_1|p_0,p_1,p_2,p_3)&=&+\rmi\,{\rm Tr}\big[{\hat\mu}_{p_3}^-(t_3){\hat\mu}_{p_1}^+(t_1)\rho_0{\hat\mu}_{p_0}^-(0){\hat\mu}_{p_2}^+(t_2)\big]\\
S_\text{ESA,RP}(T_3,T_2,T_1|p_0,p_1,p_2,p_3)&=&-\rmi\,{\rm Tr}\big[{\hat\mu}_{p_3}^-(t_3){\hat\mu}_{p_2}^+(t_2){\hat\mu}_{p_1}^+(t_1)\rho_0{\hat\mu}_{p_0}^-(0)\big],\label{eq:2DESESARP}
\end{eqnarray}
and the non-rephasing pathways are given by
\begin{eqnarray}
S_\text{GB,NR} (T_3,T_2,T_1|p_0,p_1,p_2,p_3)&=&+\rmi\,{\rm Tr}\big[{\hat\mu}_{p_3}^-(t_3){\hat\mu}_{p_2}^+(t_2){\hat\mu}_{p_1}^-(t_1){\hat\mu}_{p_0}^+(0)\rho_0\big]\\
S_\text{SE,NR} (T_3,T_2,T_1|p_0,p_1,p_2,p_3)&=&+\rmi\,{\rm Tr}\big[{\hat\mu}_{p_3}^-(t_3){\hat\mu}_{p_0}^+(0)\rho_0{\hat\mu}_{p_1}^-(t_1){\hat\mu}_{p_2}^+(t_2)\big]\\
S_\text{ESA,NR}(T_3,T_2,T_1|p_0,p_1,p_2,p_3)&=&-\rmi\,{\rm Tr}\big[{\hat\mu}_{p_3}^-(t_3){\hat\mu}_{p_2}^+(t_2){\hat\mu}_{p_0}^+(0)\rho_0{\hat\mu}_{p_1}^-(t_1)\big].\label{eq:2DESESANR}
\end{eqnarray}
When working with a sequence of laser pulses with different relative polarizations, it is necessary to adjust the electric field polarizations $p_0$, $p_1$, $p_2$, and $p_3$ accordingly.
The ESA pathways access the two-exciton manifold \cite{Cho2005,Hein2012}, which increases the number of states to propagate from $1+N_{\rm sites}$ to 
$N_{\rm states}=1+N_{\rm sites}+N_{\rm sites}(N_{\rm sites}-1)/2$
and increases the time required to compute the commutator and the bath interactions considerably.
To obtain time- and frequency-resolved two-dimensional spectra for a specific delay time $T_2=(t_2-t_1)$, the third order response function
\begin{equation}
 S(T_3=t_3-t_2,T_2,T_1=t_1)=S_{\rm RP}+S_{\rm NR}   
\end{equation}
is computed separately for the three rephasing (RP) and non-rephasing (NR) pathways. 
The resulting spectra are then Fourier transformed with different signs of $\omega_1$, according to:
\begin{eqnarray}
S_{\rm RP}(\omega_3,T_2,\omega_1)    &=&\int_0^\infty\rmd T_1 \int_0^\infty\rmd T_3\,
\rme^{-\rmi T_1\omega_1+\rmi T_3\omega_3} S_{\rm RP}(T_3,T_2,T_1)\\
S_{\rm NR}(\omega_3,T_2,\omega_1)&=&\int_0^\infty\rmd T_1 \int_0^\infty\rmd T_3\,
\rme^{+\rmi T_1\omega_1+\rmi T_3\omega_3} S_{\rm NR}(T_3,T_2,T_1).
\end{eqnarray}

\section{Ensemble averages}\label{sec:rotav}

\subsection{Isotropic average}

In typical experiments, the ensemble of randomly oriented molecules is measured with respect to the laser direction and polarization plane.
For linear spectroscopy, which records the first-order response function, rotational averaging is achieved by considering three representative electric fields along the Cartesian unit vectors:
\begin{equation}
{\bf e}_{1}=\{1,0,0\},\quad
{\bf e}_{2}=\{0,1,0\},\quad
{\bf e}_{3}=\{0,0,1\}\,.
\label{eq:laser_direction3}
\end{equation}
In two-dimensional spectra, rotational averaging becomes more involved due to the four dipole interactions involved. 
For laser pulses that share the same polarization plane, a set of ten representative electric field directions along the vertices of a dodecahedron suffices \cite{Hein2012}.

However, for more complex polarization sequences, up to twenty-one electric field combinations need to be considered.
This tensorial averaging is implemented as follows \cite{Craig1998,Gelin2017}
\begin{equation}
\langle S(T_3,T_2,T_1) \rangle_{\rm rot}=\sum_{k=1}^{3}\sum_{l=1}^{3}\sum_{m=1}^{3}\sum_{n=1}^{3} C_{klmn} S(T_3,T_2,T_1|p_{0,k},p_{1,l},p_{2,m},p_{3,n}).
\end{equation}
To perform the tensorial average, for each dipole interaction ($i=0,1,2,3$) and each pigment, a specific Cartesian component $k$ ($k=1,2,3$) of the dipole moment is selected:
\begin{eqnarray}
{\hat\mu}_{p_{i,k}}^+&=& 
\sum_{a=1}^{N_{\rm sites}}  \mathbf{e}_{k} \cdot \mathbf{d}_a |a\rangle\langle 0| \\
{\hat\mu}_{p_{i,k}}^-&=&
\sum_{a=1}^{N_{\rm sites}}  \mathbf{e}_{k}  \cdot \mathbf{d}_a |0\rangle\langle a|.
\end{eqnarray}
The factors $C_{klmn}$ are determined by
\begin{eqnarray}
C_{klmn}&=&
%(\tilde{\mu}_k\cdot\tilde{\mu}_l)(\tilde{\mu}_m\cdot\tilde{\mu}_n)
\delta_{kl}\delta_{mn}
\left[4(\mathbf{f}_0\cdot\mathbf{f}_1)(\mathbf{f}_2\cdot\mathbf{f}_3)-(\mathbf{f}_0\cdot\mathbf{f}_2)(\mathbf{f}_1\cdot\mathbf{f}_3)-(\mathbf{f}_0\cdot\mathbf{f}_3)(\mathbf{f}_1\cdot\mathbf{f}_2)\right]/30\\\nonumber
&+&
\delta_{km}\delta_{ln}
\left[4(\mathbf{f}_0\cdot\mathbf{f}_2)(\mathbf{f}_1\cdot\mathbf{f}_3)-(\mathbf{f}_0\cdot\mathbf{f}_1)(\mathbf{f}_2\cdot\mathbf{f}_3)-(\mathbf{f}_0\cdot\mathbf{f}_3)(\mathbf{f}_1\cdot\mathbf{f}_2)\right]/30\\\nonumber
&+&
\delta_{kn}\delta_{lm}
\left[4(\mathbf{f}_0\cdot\mathbf{f}_3)(\mathbf{f}_1\cdot\mathbf{f}_2)-(\mathbf{f}_0\cdot\mathbf{f}_1)(\mathbf{f}_2\cdot\mathbf{f}_3)-(\mathbf{f}_0\cdot\mathbf{f}_2)
(\mathbf{f}_1\cdot\mathbf{f}_3)\right]/30,
\end{eqnarray}
where $\mathbf{f}_i$ represents the unit vector of the electric field perpendicular to the propagation direction of the $i$th pulse $p_i$. 
Symmetry considerations reduce the $3^4=81$ $C_{klmn}$ terms to a maximum of $21$ non-zero terms, which are further reduced for specific polarization sequences.
The $C_{klmn}$ values for two exemplary polarization sequences, all parallel (all pulses have the same polarization), and double-crossed (each pair of pulses has the polarization rotated by $90^\circ$) are listed in Table~\ref{tab:Cklmn}.

\begin{table*}[bt]
\begin{center}
{\footnotesize
\begin{tabular}{cc}
$(k,l,m,n)$ & $C_{klmn}$ $\langle 0^\circ ,0^\circ, 0^\circ ,0^\circ \rangle$ \\\hline
$(1, 1, 2, 2)$, $(1, 1, 3, 3)$, $(1, 2, 1, 2)$, $(1, 2, 2, 1)$, $(1, 3, 1, 3)$, $(1, 3, 3, 1)$ & $+\frac{1}{15}$ \\ 
$(2, 1, 1, 2)$, $(2, 1, 2, 1)$, $(2, 2, 1, 1)$, $(2, 2, 3, 3)$, $(2, 3, 2, 3)$, $(2, 3, 3, 2)$ & $+\frac{1}{15}$ \\
$(3, 1, 1, 3)$, $(3, 1, 3, 1)$, $(3, 2, 2, 3)$, $(3, 2, 3, 2)$, $(3, 3, 1, 1)$, $(3, 3, 2, 2)$ & $+\frac{1}{15}$ \\
$(1, 1, 1, 1)$, $(2, 2, 2, 2)$, $(3, 3, 3, 3)$ & $+\frac{1}{5}$ \\
& \\
$(k,l,m,n)$ & $C_{klmn}$ $\langle 45^\circ ,-45^\circ, 90^\circ ,0^\circ \rangle$ \\\hline
$(1, 2, 1, 2)$, $(1, 2, 2, 1)$, $(1, 3, 1, 3)$ &  $+\frac{1}{12}$,$-\frac{1}{12}$,$+\frac{1}{12}$ \\
$(1, 3, 3, 1)$, $(2, 1, 1, 2)$, $(2, 1, 2, 1)$ & $-\frac{1}{12}$,$-\frac{1}{12}$,$+\frac{1}{12}$ \\
$(2, 3, 2, 3)$, $(2, 3, 3, 2)$, $(3, 1, 1, 3)$ & $+\frac{1}{12}$,$-\frac{1}{12}$,$-\frac{1}{12}$ \\
$(3, 1, 3, 1)$, $(3, 2, 2, 3)$, $(3, 2, 3, 2)$ & $+\frac{1}{12}$,$-\frac{1}{12}$,$+\frac{1}{12}$
\end{tabular}
}
\end{center}
\caption{$C_{klmn}$ coefficients for isotropic averaging of the $\langle 0^\circ ,0^\circ, 0^\circ ,0^\circ \rangle$ and $\langle 45^\circ ,-45^\circ, 90^\circ ,0^\circ \rangle$ polarization sequences
}\label{tab:Cklmn}
\end{table*}

\subsection{Static disorder}

A second type of averaging is required to account for variations in site energies in the molecular ensemble, for instance caused by slow movement/bending of the complex.
The resulting variations of the site energies (inter-site couplings are typically less affected by those modes), requires to average various realizations of the excitonic Hamiltonian.
This variation is termed \textit{static disorder}.
In the simplest case, for the absorbtion spectra of a monomeric unit, this leads to an additional broadening of the spectral line-shapes in addition to the thermal line-shape function discussed before.
For more complex spectra, such as 2DES, the inclusion of static disorder has non-trivial effects on various locations of the 2DES \cite{kramer_effect_2020}.

\section{Using HEOM for quantum dynamics}\label{sec:heom}

The hierarchical equations of motions require to propagate a large (up to $10^6$) number of interlinked matrices in parallel.
A computationally efficient HEOM implementation \cite{Kreisbeck2011,Strumpfer2012a,Kramer2018e,Noack2018a} distributes the computations across many threads and benefits from many-core processors (either many-core CPUs or GPUs).
The \textit{distributed memory} DM-HEOM \cite{Kramer2018e,Noack2018a} tool provides a comprehensive set of \textit{applications} to compute the time-evolution of a density matrix, linear spectra, and two-dimensional spectra.
DM-HEOM is distributed as C++/OpenCL source code \cite{noack_dm-heom_2023}.
A ready-to-run implementation of HEOM using GPUs is provided on the \textit{https://nanohub.org} simulation platform \cite{Kreisbeck2014}.

\subsection{Populations and coherences}

The spectral density sets the time-scale for the duration of excitonic coherences, as discussed in Sect.~\ref{sec:dec}. 
Using HEOM, the conditions for long-lived electronic coherences in the FMO photosynthetic complex have been investigated in \cite{Kreisbeck2012,Kreisbeck2013}.
To compute the overall efficiency and transport-time through a network of coupled chromophores requires to additionally consider loss channels, i.e.\ due to radiative decay of the excitons.
This has been explored in \cite{Kreisbeck2011,Kramer2014} and shows that from the theoretical models an intermediate coupling to the bath is preferred to facilitate a fast thermalization (see also Sect.~\ref{sec:dec}).

\subsection{Two-dimensional spectra}

\begin{figure}
    \centering
    \includegraphics[width=0.5\textwidth]{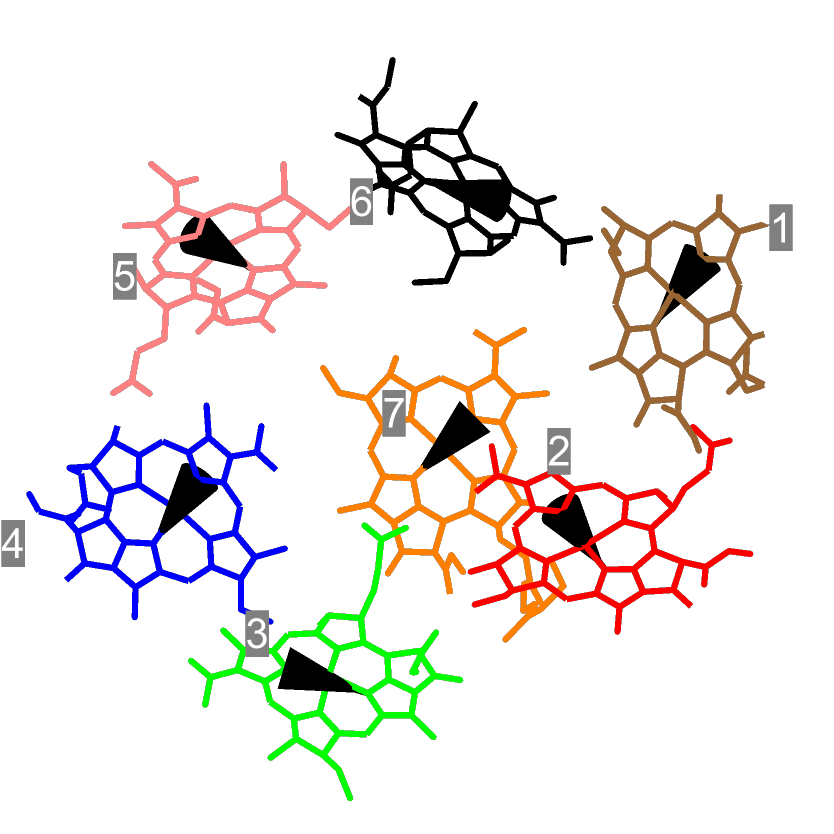}
    \caption{Monomeric unit of the FMO complex with 7 bacteriochlorophylls with arrows indicating the directions of the transition dipoles. The protein scaffold keeping the  bacteriochlorophylls in place is not shown.
    }
    \label{fig:fmodip}
\end{figure}

The simulation of 2DES signals using Eqs.~(\ref{eq:2DESGBRP}-\ref{eq:2DESESANR}) and HEOM proceeds by computing the time-evolution of the reduced density matrix and the application of the dipole operator at the specific times $t_1,t_2,t_3$.
Initially the reduced density matrix at $t=0$ represents a populated  ground state 
$\rho(0)=\sigma_0(0)=|0\rangle\langle 0|$.
In addition to the exciton Hamiltonian, also the relative transition dipole strengths and directions are required as input parameters.
For the FMO complex, the transition dipoles are directed along the nitrogen atoms NB-ND in the molecular structure (PDB:3ENI) \cite{Adolphs2008}. See also Tab.~1 in \cite{Kramer2018e} and Fig.~\ref{fig:fmodip}

\begin{figure}
    \centering
    \includegraphics[width=0.32\textwidth]{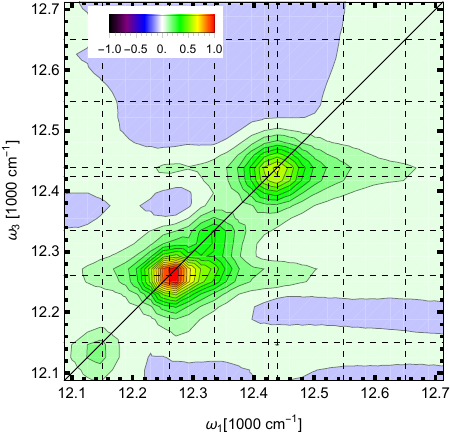}
    \includegraphics[width=0.32\textwidth]{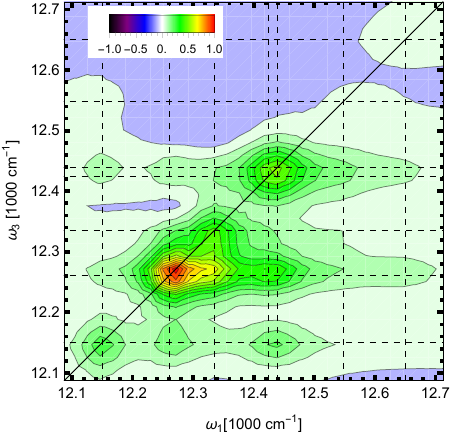}
    \includegraphics[width=0.32\textwidth]{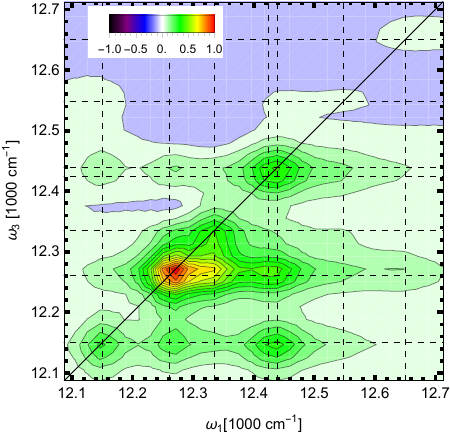}\\[2ex]
    \includegraphics[width=0.32\textwidth]{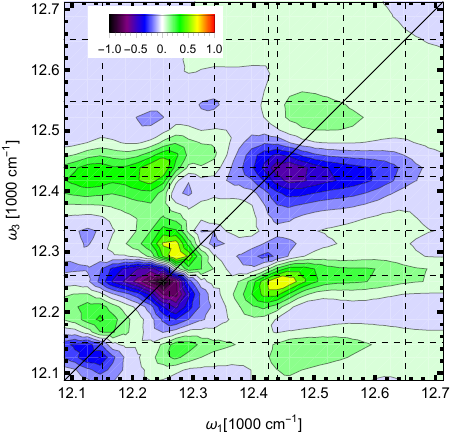}
    \includegraphics[width=0.32\textwidth]{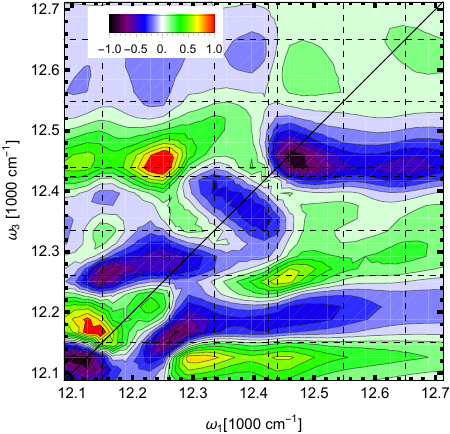}
    \includegraphics[width=0.32\textwidth]{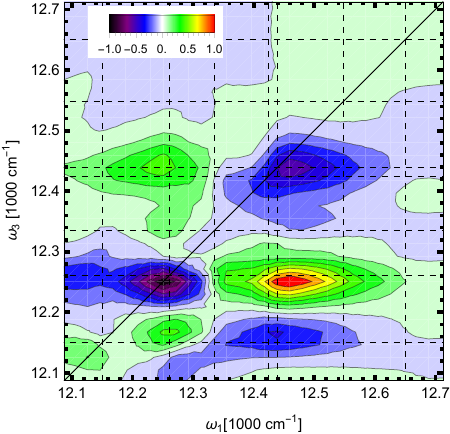}
    \caption{2DES (sum of rephasing and non-rephasing pathways) for the FMO complex computed with DM-HEOM from left panels to right panels for increasing delay time $T_2=\{40,400,800\}$~fs at temperature $100$~K. 
    Upper row: all parallel polarization $\langle 0^\circ ,0^\circ, 0^\circ ,0^\circ \rangle$. Lower row: double-crossed polarization $\langle 45^\circ ,-45^\circ, 90^\circ ,0^\circ \rangle$. 
    Rotational averaging is performed, static disorder is not considered.}
    \label{fig:2dfmo}
\end{figure}

Typical 2DES of the Fenna-Matthews-Olson (FMO) complex for the $\langle 0^\circ ,0^\circ, 0^\circ ,0^\circ \rangle$ polarization sequence computed using the DM-HEOM method \cite{Noack2018a, Kramer2018e} are presented in Figure~\ref{fig:2dfmo}, upper row.
The Hamiltonian and dipole directions used in the computations can be found in Table 1 and Equation (77) of Kramer et al. \cite{Kramer2018e}, respectively.
The parameter file for DM-HEOM \cite{Noack2018a,Kramer2018e,noack_dm-heom_2023} to generate the upper right panel reads:
\begin{lstlisting}[language={Ini},breaklines=true]
[program]
task=two_dimensional_spectra
observations={(matrix_trace_two_dimensional_spectra, fmo_0000_400fs.dat)}
observe_steps=2

[filtering]
strategy=none
first_layer=-1

[solver]
stepper_type=rk_rk4
step_size=4.e-15
steps=2
track_flows=false
flow_filename=

[system]
ado_depth=3
sites=7
hamiltonian={{1410.000, -87.70000, 5.500000, -5.900000, 6.700000, -13.70000, -9.900000}, {-87.70000, 1530.000, 30.80000, 8.200000, 0.7000000, 11.80000, 4.300000}, {5.500000, 30.80000, 1210.000, -53.50000, -2.200000, -9.600000, 6.000000}, {-5.900000, 8.200000, -53.50000, 1320.000, -70.70000, -17.00000, -63.30000}, {6.700000, 0.7000000, -2.200000, -70.70000, 1480.000, 81.10000, -1.300000}, {-13.70000, 11.80000, -9.600000, -17.00000, 81.10000, 1630.000, 39.70000}, {-9.900000, 4.300000, 6.000000, -63.30000, -1.300000, 39.70000, 1440.000}}

[baths]
max_per_site=1
number=7
coupling={{0}, {1}, {2}, {3}, {4}, {5}, {6}}
lambda={35, 35, 35, 35, 35, 35, 35}
invnu={50, 50, 50, 50, 50, 50, 50}
Omega={0, 0, 0, 0, 0, 0, 0}
matsubaras=1
temperature=100

[dipole]
directions={{0.74101,0.56060,0.36964}, {0.85714,-0.50378,0.10733}, {0.19712,-0.95741,0.21097}, {0.79924,0.53357,0.27661}, {0.73693,-0.65576,-0.16406}, {0.13502,0.87922,-0.45689}, {0.49511,0.70834,0.50310}}
centers={{-0.7410,-0.5606,-0.3696}, {-0.8571,0.5038,-0.1073}, {-0.1971,0.9574,-0.2110}, {-0.7992,-0.5336,-0.2766}, {-0.7369,0.6558,0.1641}, {-0.1350,-0.8792,0.4569}, {-0.4951,-0.7083,-0.5031}}
strengths={1,1,1,1,1,1,1}
tensor_prefactors={0.2, 0.066667, 0.066667, 0.066667, 0.066667, 0.066667,0.066667, 0.066667, 0.066667, 0.066667, 0.2,0.066667, 0.066667, 0.066667, 0.066667, 0.066667, 0.066667, 0.066667, 0.066667, 0.066667, 0.2}
tensor_components={{0,0,0,0}, {0,0,1,1}, {0,0,2,2}, {0,1,0,1}, {0,1,1,0}, {0,2,0,2},{0,2,2,0}, {1,0,0,1}, {1,0,1,0}, {1,1,0,0}, {1,1,1,1}, {1,1,2,2}, {1,2,1,2},{1,2,2,1}, {2,0,0,2}, {2,0,2,0}, {2,1,1,2}, {2,1,2,1}, {2,2,0,0}, {2,2,1,1},{2,2,2,2}}

[spectra]
steps_t_1=200
steps_t_3=200
steps_t_delay=100
pathways={gbnr,senr,esanr,gbrp,serp,esarp}
\end{lstlisting}
The resulting file contains the third order non-linear response function $S(T_1,T_2,T_3)$ for $T_1, T_3=0\ldots 200\times4\times10^{-15}$~s and delay time $T_2=100\times4\times10^{-15}$~s (all steps are measured in multiples of \verb+step_size+).
In this example a total of $6 \times 21$ spectra are computed to perform the isotropic ensemble average for all six rephasing and non-rephasing pathways.
The energy transfer process shown is revealed from left to right in the upper panels in Fig.~\ref{fig:2dfmo} by the appearance of off-diagonal peaks below the diagonal line: a lower cross-peak implies a reduced probe (de-excitation) energy as compared to the initial excitation energy: energy has been dissipated to the vibrational modes. 
The dissipation drives a directional energy flow towards thermal equilibrium which implies a higher occupation of lower lying energy states.

This energy transfer is clearly observable in experimental measurements performed by Brixner et al.\ \cite{Brixner2005}, see also the HEOM computations by Hein et al.\ \cite{Hein2012} and Kreisbeck et al. \cite{Kreisbeck2012}.
Computed spectra can be separated into contributions from ground state bleaching, stimulated emission, and excited state absorption (see Fig.~3 in \cite{Kramer2017a} for an illustrative example).
The energy flow is not reflected by the ground state bleaching signal, since it requires a population transfer.

In addition to energy decay, Engel et al.\ \cite{Engel2007} and Panitchayangkoon et al.\ \cite{Panitchayangkoon2010} reported the presence of oscillatory amplitudes in the 2DES signals. 
These oscillations are attributed to a combination of ground-state bleach-induced vibrational modes and electronic coherences.
In computed 2DES these contributions to oscillatory signals can be cleanly separated by a short time Fourier transform \cite{Kreisbeck2013}.
The electronic coherences are expected to decay on a timescale determined by the combined dephasing and relaxation decoherence time \cite{Kreisbeck2012} of the two eigenenergies at the location of the cross-peak. 
In addition to the electronic coherences vibrational peaks in the spectral density are present in 2DES signals, in particular in the ground sate bleaching part.
These contribution can persist longer than the electronic coherences and overshadow them \cite{Kreisbeck2012,Kreisbeck2013}.
The pure dephasing time is influenced by the slope of the spectral density $J(\omega)$ towards zero frequency, while the relaxation rate is determined by the value of the spectral density at the eigenenergies.
Both factors contribute to the decay time, as shown in the supplementary information of Kreisbeck et al.\ \cite{Kreisbeck2012} and Fig.~\ref{fig:coherence}.

The reorganization energy $\lambda_m$, which is related to the spectral density $J_m(\omega)$ of each pigment $m$, manifests as a downward shift of the diagonal and cross peaks as the delay increases. This shift is consistent with the reorganization energies of approximately $40$~cm$^{-1}$ assigned to the bacteriochlorophylls in the FMO complex by Adolphs et al. \cite{Adolphs2006}.

Experimental data for different polarization sequences for the FMO complex is presented in \cite{Thyrhaug2016,Thyrhaug2018} and requires corresponding theoretical models for interpretation.
While in the all parallel polarization sequence ($\langle 0^\circ ,0^\circ, 0^\circ ,0^\circ \rangle$), 
all isotropic averaging coefficients are positive numbers, for other polarizations cancellation effects due to alternating signs (see Tab.~\ref{tab:Cklmn}) occur.
These lead to additional structures in the 2DES \cite{kramer_effect_2020}, and Fig.~\ref{fig:2dfmo}, lower row.

To facilitate analysis and for studying various static disorder configuration, an efficient storage and interpolation of 2DES results is useful. 
A highly compressed storage uses custom neural networks \cite{Rodriguez2019b}.
Once the neural network has been trained with an extensive date set of exemplary computations, it generates a 2DES for a specific set of site energies of the Hamiltonian.
This approach has been used to simulate 2DES data sets for differently prescribed static disorder distributions for the FMO complex \cite{kramer_effect_2020}.

\section{Summary.}

The open quantum system approach provides the required tools for tracking the energy flow in molecular complexes.
Only for small or large couplings simplified descriptions of the density matrix are available, while in general more accurate quantum propagation methods are required, such as HEOM.
These methods come with an additional computational overhead, which requires to use highly optimized numerical algorithms.
Compared to the numerical effort to obtain a linear absorption spectra, 2DES calculations increase the run times by $\sim 10^4$ due to the complexity of ensemble averaging and the need to span three time intervals.
The interpretation and analysis of 2DES needs theoretical models to assess the impact of the different pathways and ensemble averages.

\paragraph*{Acknowledgments.}

The further development of numerical methods for describing open quantum systems is funded by the Austrian Science Fund (FWF project ``Open quantum dynamics lab'') and by the NVIDIA academic hardware donation grant ``Simulating branched flow with tensor processing units''.

\printbibliography

\end{document}